\documentclass[reprint,aip,amsmath,amssymb,nofootinbib]{revtex4-2}
\usepackage{amssymb}
\usepackage{amssymb}
\usepackage{amssymb}
\usepackage{graphicx}
\usepackage{graphics}
\usepackage{amsmath}
\usepackage{placeins}
\usepackage{hyperref}
\usepackage[dvipsnames]{xcolor}
\usepackage{wasysym}
\usepackage{soul}
\usepackage{float}
\usepackage{array}
\usepackage{tabulary}
\usepackage{caption}
\usepackage{fancyhdr}
\usepackage{wrapfig}

\usepackage[normalem]{ulem}
\usepackage{graphicx}
\usepackage{color,soul}
\usepackage{subfigure}
\usepackage{dcolumn}
\usepackage{bm}
\usepackage{hyperref}
\usepackage{url}
\usepackage{multirow}
\hypersetup{
    colorlinks=true,
    linkcolor=purple,
    filecolor=blue,      
    urlcolor=blue,
    citecolor=blue,
}
\usepackage{float}

\usepackage[makeroom]{cancel}

\begin{document}

\title{Fast Pattern Recognition for Electron Emission Micrograph Analysis}

    \author{\firstname{Taha Y.} \surname{Posos}}
	\email{posostah@msu.edu}
    \affiliation{Department of Electrical and Computer Engineering, Michigan State University, MI 48824, USA}
    \author{\firstname{Oksana} \surname{Chubenko}}
    \affiliation{Department of Physics, Arizona State University, AZ 85281, USA}
	\author{\firstname{Sergey V.} \surname{Baryshev}}
	\email{serbar@msu.edu}
	\affiliation{Department of Electrical and Computer Engineering, Michigan State University, MI 48824, USA}

\begin{abstract}
	In this work, a pattern recognition algorithm was developed to process and analyze electron emission micrographs. Various examples of dc and rf emission are given that demonstrate this algorithm applicability to determine emitters spatial location and distribution and calculate apparent emission area. The algorithm is fast and only takes $\sim$10 seconds to process and analyze one micrograph using resources of an Intel Core i5.
\end{abstract}

\maketitle

\section{Introduction}\label{S:intro}
Many studies\cite{locally,mypaper,Jiahang1,Jiahang2,cui1,kolosko} have convincingly demonstrated that electron emission from a large surface area field emission cathode placed in a macroscopic electric field $E$ is not uniform. The area contributing to emission is only a small portion of the total surface area available for emission. As it appears on the imaging screen, most of the emission is confined to a small number of emission spots randomly distributed over the surface. Therefore, a proper and thorough estimation of the apparent emission area and a number of emission locations from  micrographs obtained in imaging experiments is essential to quantify emitters in terms of the current density $j$ and its spatial variation, as well as to establish an accurate relation between $j$ and $E$-field. The importance of developing such methodologies is two-fold. One is practical: an actual emission area needs to be known to compare cathode materials produced by various or varied syntheses. Second is fundamental: only a properly established $j$--$E$ (and not $I$--$V$) relationship can help define the validity range of a classical Fowler-Nordheim (FN) emission and clarify the role of other mechanisms at play that cause deviation from the FN emission, i.e. cause non-conventional behavior, that is have being observed across a large body of experimental work\cite{locally,mypaper,Jiahang1,Jiahang2,cui1,kolosko,serbun2013,GaN,KOCK,Obraz}.

A simple and convenient way to measure distribution of electron emission sites, and thus the emission area, is using a luminescence (or phosphor) screen, also known as a scintillator. These screens emit light when interacted with electrons. When such a screen is used as an anode in a field emission experiment, it will magnify and project an emission pattern formed on the cathode surface under the external field force. When captured by a camera, it creates a micrograph. An experimental system can be designed such that $I$--$V$ curves can be taken synchronously with micrographs. This, in turn, provides for the field emission current and apparent emission area measurements. This work is motivated by the lack of a guided micrograph processing for emission area calculations, and a fast processing/analysis algorithm is proposed. Results are presented for micrographs obtained from ultra-nano-crystalline dimond (UNCD) films and carbon nano-tubes (CNTs) under dc field using screens made of Ce-doped yttrium aluminum garnet (YAG:Ce), which produces a bright green luminescence line at 550 nm. Nevertheless, the proposed image processing algorithm is applicable to any other phosphor screen. Additional applicability for pattern recognition of field emission micrographs obtained under pulsed rf conditions is demonstrated.

The algorithm is realized on the MATLAB platform and takes an advantage of its strength in processing of arrayed data. When compared to an earlier version implemented in Mathematica,\cite{locally} where a server was required to process extensive sets of micrographs, our present implementation performs 10 times faster on a personal laptop. The image processing code is an open-source software: examples and the user manual are being prepared. The first and future releases can be found on \href{https://github.com/schne525}{GitHub}.

\section{Feature Extraction}\label{S:extraction}

\begin{figure*}
	\centering\includegraphics[scale=0.23]{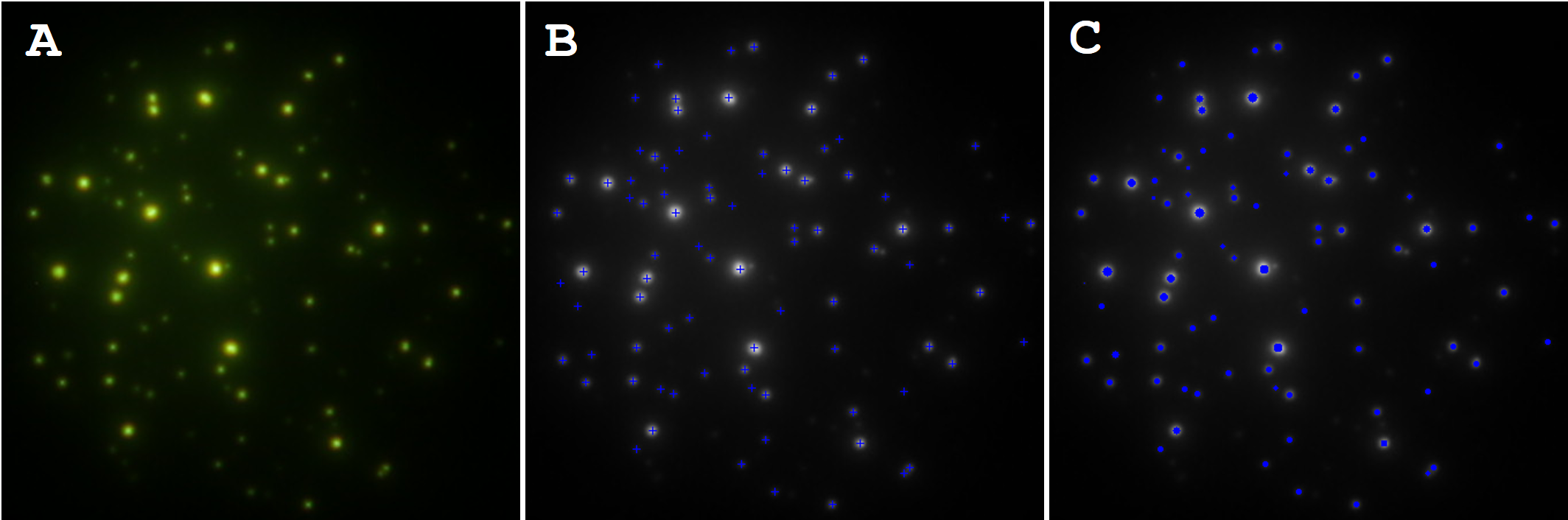}
	\caption{A) A typical micrograph obtained from an UNCD film under the applied dc field.\cite{locally} A 450$\times$450 px$^2$ image represents a projection of spatial distribution of electron emission sites onto a YAG anode placed 106 $\mu$m from a 4.4 mm diameter cathode. B) Detected LMs (shown with blue plus signs) overlaid with emission spots shown on the micrograph. C) Emission pixels (shown in blue), representing the projected emission area, overlaid with emission spots shown on the micrograph.}\label{F:uncdimg}
\end{figure*}

\begin{figure*}
	\centering\includegraphics[scale=0.23]{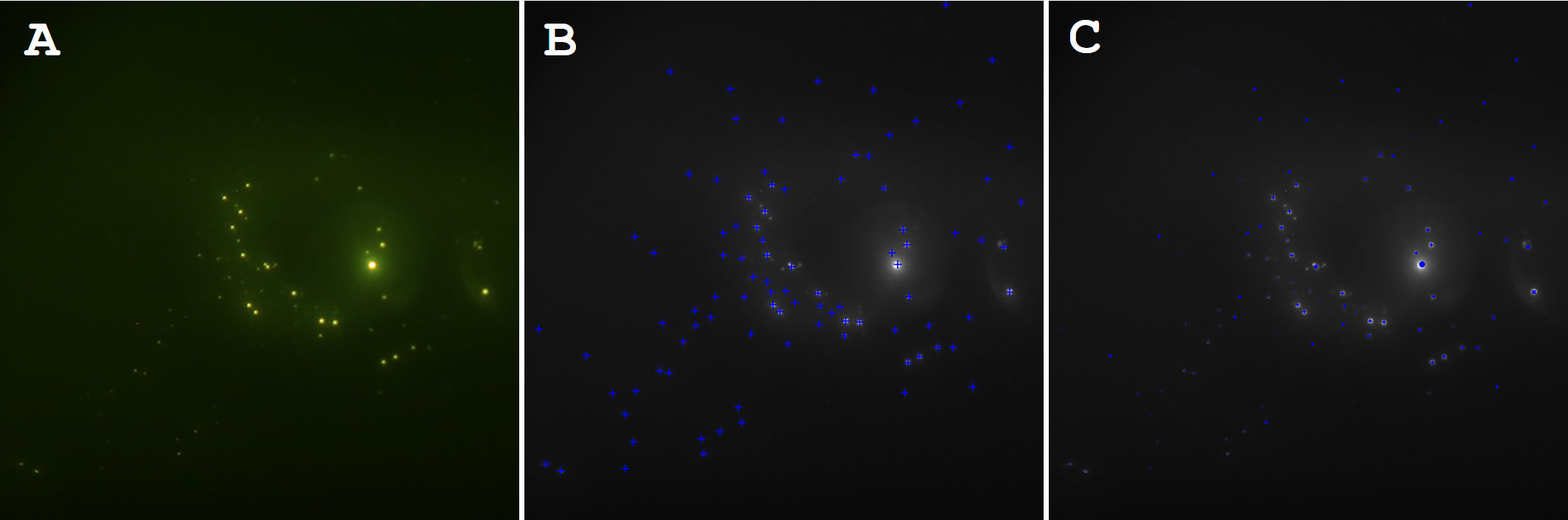}
\caption{A) A typical micrograph obtained from a CNT fiber under dc field. B) Detected LMs (shown with blue plus signs) overlaid with emission spots shown on the micrograph. C) Emission pixels (shown in blue), representing the projected emission area, overlaid with emission spots shown on the micrograph.}\label{F:cntimg}
\end{figure*}

As shown in the examples below, emission patterns can vary a lot, with an additional challenge being a bright halo background, which complicates image analysis. However, when emission patterns under dc\cite{locally,mypaper,cui1} and rf\cite{Jiahang1,Jiahang2} fields are compared (see Figs.~\ref{F:uncdimg}A, \ref{F:cntimg}A, \ref{F:cnt8}A, and \ref{F:uncdrfimg}A), one thing is found to be in common: the patterns consist of bright spots or streaks separated spatially enough to distinguish them visually. The strategy is to detect the brightest pixel within each emission spot, which is further referred to as a local maximum (LM). Then the number of electron emission sites is equal to the LM count and the emission area can be estimated by assembling certain neighbor pixels around LMs.

Typical image is in 8-bit gray scale format, and pixel value is between 0 and 255. If an image is in 24-bit RGB format, it is converted into gray scale by averaging out red, green and blue pixel values.

\begin{figure}[!b]
	\centering\includegraphics[scale=0.40]{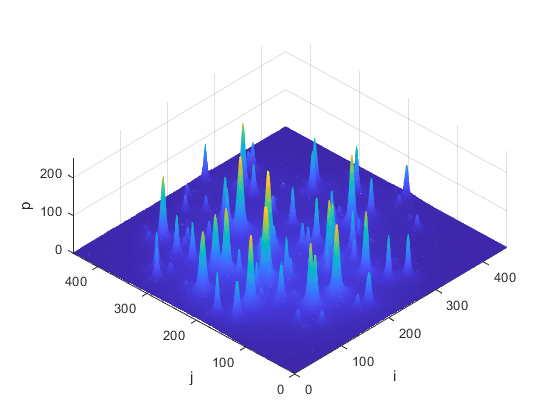}
	\caption{A 3D plot of a micrograph shown in Fig.~\ref{F:uncdimg}A.}\label{F:3dgauss}
\end{figure}

When a typical micrograph (Fig.~\ref{F:uncdimg}A) is represented as a 3D plot (Fig.~\ref{F:3dgauss}), it can be seen that emissions spots appear as Gaussian peaks atop a certain background. LMs are the brightest pixels of each Gaussian peak and they are also well separated in space. Therefore, two features must be known for each pixel in order to define LMs: pixel value and the distance to the nearest brighter pixel. Because the pixel values is an integer, occurrence of more than one LM for each peak is possible. In order to prevent this, small random noise between 0-0.1 in value, excluding 0 and 0.1, is added to the image so that there are no pixel values equal to another one. This procedure does not perturb the image because the original image can be retrieved by rounding each pixel value.

A pixel $ a $ with spatial coordinates $ (i_{a},j_{a})$, where $ i_{a} $ is the row number and $ j_{a} $ is the column number in a 2D array of digitized image, will be represented by $ (p_{a},d_{a}) $ in a feature space, where $ p_{a} $ is the intensity feature and $ d_{a} $ is the distance feature. $p_{a}$ is just a pixel value (an integer, typically between 0 and 255), which can be simply extracted from a data array (before adding the noise).

A fast method of extracting the distance feature is searching for a brighter neighbor in a certain neighborhood of each pixel, which is called the search region\cite{peaksearch, maxfilter}. The search region must be large enough to enclose an entire Gaussian-like peak, but small enough to enclose not more than one entire Gaussian peak. All images presented in this work were 450$\times$450 px$^2$. Although the size of emission spots varied, two standard deviations of each Gauss peak corresponded to 10 pixels or less. At the same time, the distance between peaks was 20 pixels or more. Therefore, a circular search region was chosen: centered around each pixel, the search region radius was set to 10 pixels. It should be noted that for different images or image sizes, this value may have to be adjusted accordingly. The Euclidean distance between any two pixels $a'$ at $(i_{a'}, j_{a'})$ and $a$ at $(i_a, j_a)$ in the image is defined as:

\begin{equation}\label{E:distance}
d_{a'a}=\sqrt{\left( i_{a'}-i_a\right)^{2}+\left( j_{a'}-j_a\right)^{2} }.
\end{equation}
If $s_a$ is the search region for a pixel $ a $ carrying value $p_a$, and a pixel $ b $ carrying value $p_b$ contained inside the $s_a$ is the next closest pixel to the pixel $ a $  such that $p_b>p_a$, then the distance $d_a$ assigned to the pixel $ a $ is
\begin{equation}\label{E:normaldistance}
d_a=\sqrt{\left( i_a-i_b\right)^{2}+\left( j_a-j_b\right)^{2} }.
\end{equation}
\noindent If no other brighter pixel was found in a search region, the brightest pixel is called the maximum in search region (MISR). There is a distance property that can be defined for a MISR as the distance to another closest MISR that is brighter than the former. Say, pixel $A$ with value of $p_A$ is the MISR in its own search region $s_A$, and pixel $B$ with value of $p_B$ is the MISR in its own search region $s_B$, and $p_B > p_A$ while $p_B$ is a MISR closest to pixel $A$, then the distance $d_A$ assigned to pixel $A$ reads
\begin{equation}\label{E:lmdistance}
d_A = \sqrt{\left( i_A-i_B\right)^{2}+\left( j_A-j_B\right)^{2} }.
\end{equation}
The brightest pixel across the entire image is the global maximum (GM). There is a distance value that is assigned to the GM. It is the distance to closest MISR (regardless of its value). Consider, pixel $ GM $ with value of $p_{GM}$ is the global maximum and pixel $ A $ with value of $p_A $ is a MISR, closest to $ GM $, the distance assigned to the $ GM $ is
\begin{equation}\label{E:globaldistance}
d_{GM} = \sqrt{\left( i_{GM}-i_A\right)^{2}+\left( j_{GM}-j_A\right)^{2}}.
\end{equation}

\begin{figure*}
	\centering\includegraphics[scale=0.23]{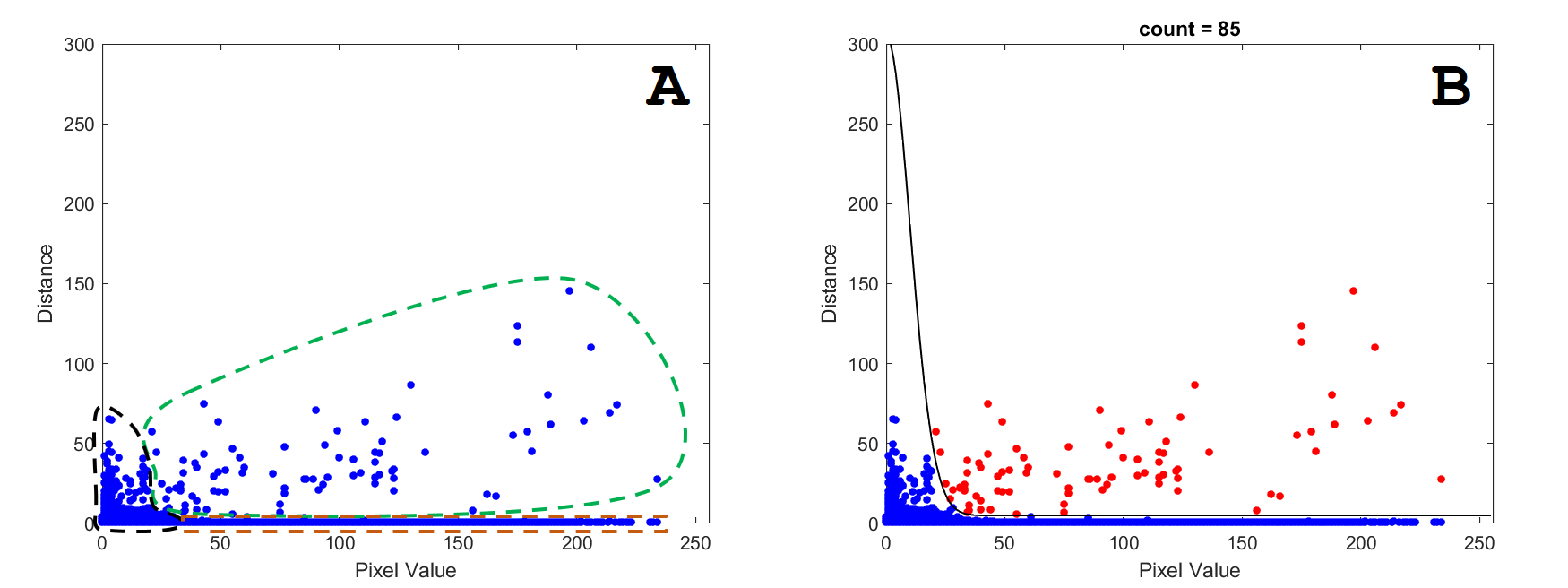}
\caption{A) Decision plot of extracted features for the micrograph obtained from an UNCD cathode operated under the dc field. Green, black and brown dashed regions show where LMs, background pixels, and emission pixels are mostly located in the decision plot, respectively. There is no overlapping between an LM cluster and a uniform background. B) A black curve shows the Gaussian decision boundary. The pixels shown in red are classified as LMs.}\label{F:uncddb}
\end{figure*}

\begin{figure*}
	\centering\includegraphics[scale=0.23]{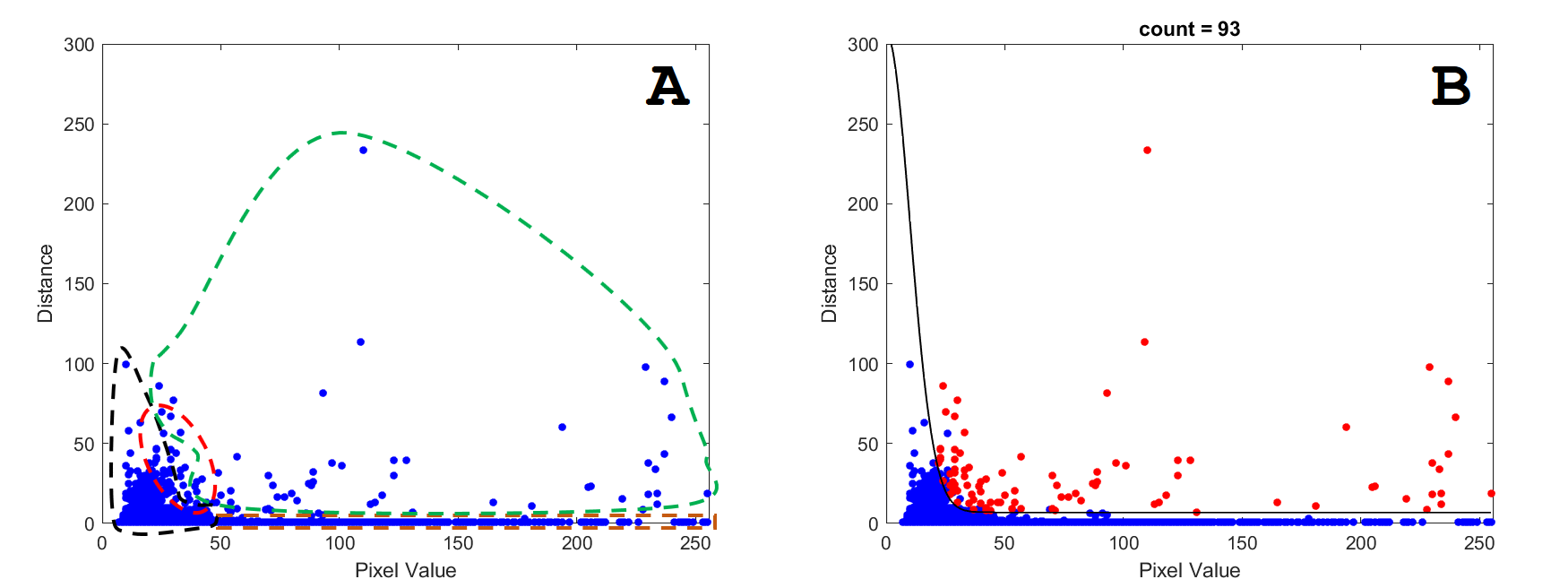}
	\caption{A) Decision plot of extracted features for the micrograph obtained from a CNT fiber under the dc field. Green, black, and brown dashed regions show the regions where LMs, background pixels, and emission pixel lie roughly. The red dashed region shows overlap between LMs and background cluster due to gradient and intensity of the background in Fig.~\ref{F:cntimg}. B) Black curve is an applied Gaussian decision boundary. Red points are detected LMs. Notice that not all points above the boundary are red; this is because some false LMs were filtered out.}\label{F:cntdb}
\end{figure*}

After all these distances are extracted and recorded, the random noise added earlier gets removed. The code stores both $(p_a,d_a)$ and $ (i_{a},j_{a}) $ pairs for every pixel. Scatter plots of intensity--distance arrays, which correspond to emission micrographs from UNCD (Fig.~\ref{F:uncdimg}A) and CNTs (Fig.~\ref{F:cntimg}A), are presented in Figs.~\ref{F:uncddb}A and \ref{F:cntdb}A, respectively. Such plots are called $decision$ $plots$, also known as $plot$ $of$ $features$ in machine learning literature.

\section{Decision Boundary}\label{S:DB}

Backlight/background gradient, and pixel values and density -- all continuously varying as $I$--$V$ curves are measured -- prompt MISRs and LMs to differ. Consider two emission spots next to each other: there might be more than one peak in a search region (a few instances can be seen in Fig.~\ref{F:3dgauss}). It means that one spot (fainter among the two) will not be on the the MISR list, although it is a LM. Consider another: there is a fairly faint but distinct emission spot atop a background that has steep gradient nature. Therefore, a background pixel may be found brightest in a search region for that faint emission spot, and consequently would get listed as a MISR, although it was not a LM. As a result, some MISRs are not necessarily LMs. The opposite is also true -- some LMs are not necessarily MISRs. Therefore, LMs need to be detected from the decision plot by estimating an appropriate decision boundary.

A supervised machine learning using labeled data cannot be used for this class of problems because labeling hundreds of images with hundreds of emission spots is not feasible. In addition, micrographs vary a lot for different materials, geometries, or evolve with a high dynamic range throughout a single experiment. An unsupervised machine learning scheme might be used. However, as seen from Figs.~\ref{F:uncddb}A and \ref{F:cntdb}A, no distinct clusters form on the decision plots. A simple robust approach using a tunable decision boundary of a certain form is attempted here.

For pixels that come from a continuous low-level background (intense and high-gradient background case will be discussed later), the algorithm finds a brighter pixel less than a few pixels away. Thus, these background pixels are characterized by a small distance $d_a$ and a small intensity $p_a$.
As a rule, the background pixels outnumber the emission spot pixels. Therefore, most of the background pixels lie near origin and appear as a crowded cluster encircled by the black dashed loop in Figs.~\ref{F:uncddb}A and \ref{F:cntdb}A. Some pixels that belong to emission spots are also characterized by a small distance (less than the radius of the search region), but most of them are brighter than the background and their pixel values are large, up to 255. Therefore, these pixels form a stretched cluster at the bottom of the decision plot as shown by the brown dashed region in Figs.~\ref{F:uncddb}A and \ref{F:cntdb}A.

LMs are bright pixels separated from brighter pixels by a large distance (usually larger than the search radius). They are shown by dashed green regions in decision plots. If the background is low and uniform (as in the case shown in Fig.~\ref{F:uncdimg}A), LMs can be easily separated from background pixels (see the green dashed region in a corresponding decision plot in Fig.~\ref{F:uncddb}A). However, when a stronger background appears with a distinct gradient across the image plane as shown in Fig.~\ref{F:cntimg}A, there are relatively faint emission pixels that are barely noticeable on a bright background. In Fig.~\ref{F:cntdb}A, these pixels are shown inside the red dashed region and are intermixed with background pixels. To separate LMs from the background, we use a Gaussian decision boundary given by
\begin{equation}\label{E:gauss}
f(p_a)=A\cdot e^{-\frac{(p_a-\mu)^{2}}{2\sigma^{2}}} + k,
\end{equation}
where $ A $ is the amplitude, $ \mu $ is the mean, $ \sigma $ is the standard deviation, and $ k $ is an offset. Functions, which define decision boundaries for micrographs in Figs.~\ref{F:uncdimg}A and \ref{F:cntimg}A, are shown in Figs.~\ref{F:uncddb}B and \ref{F:cntdb}B, respectively. The decision rule is as follows. Consider a pixel $ a $ carrying the pair of values $ (p_a,d_a) $. If $ d_a>f(p_a) $, then pixel $ a $ is a LM. The parameters $A$, $\mu$, $\sigma$, and $k$ can be adjusted for each data set. If the background and LM pixels are mixed as shown by the red dashed region in Fig.~\ref{F:cntdb}A, many background pixels are detected as LMs. One way to fix it is to increase $ \sigma $ to exclude background pixels from the LM list. In a case when the density of emission spots is high (emission spots, and thus LMs, are spatially very close to each other), decreasing $ k $ helps to identify some missing LMs. The distance feature $d_a$ was never greater than 300 for 450$\times$450 px$^2$ images. Thus, the parameter $A$ was kept at the value of 300 for all data sets presented. Mean $ \mu $ was set to zero for all data sets.

Any false LM can be further filtered out by applying the surface fitting method given in Sec.~\ref{S:area}. Therefore, a crude adjustment of the decision boundary parameters is enough.

Decision boundaries shown with black curves in Fig.~\ref{F:uncddb}B and Fig.~\ref{F:cntdb}B were applied to Fig.~\ref{F:uncddb}A and Fig.~\ref{F:cntdb}A, respectively. Then, false LMs were filtered out by the curve fitting method. A final list of detected LMs are labeled with red data points on the decisions plots in Figs.~\ref{F:uncddb}B and \ref{F:cntdb}B as well as with blue crosses in Figs.~\ref{F:uncdimg}B and \ref{F:cntimg}B. These examples illustrate nearly perfect LM detection.

\section{Emission Area}\label{S:area}

As it was already mentioned, each emission spot on a micrograph (e.g., Fig.~\ref{F:uncdimg}A) appears on a 3D plot (Fig.~\ref{F:3dgauss}) as a quasi-symmetric Gaussian with its center at a LM. Once a LM is detected, a 2D Gaussian function can be used to fit an intensity peak\cite{mpfit, blobdetection}. The fitting function, which returns the estimated pixel value $ p_{ij}^{e} $ for a pixel at the position $ (i,j) $, is given by 
\begin{equation}\label{E:gaussfit}
p_{ij}^{e}=A\cdot e^{-\frac{(i-i_{LM})^2+(j-j_{LM})^2}{2\sigma^2}}+C,
\end{equation}
where $ A $ is the amplitude, $ \sigma $ is the standard deviation, $C$ is an offset from an $i$--$j$ plane to manage the background level, and $(i_{LM},j_{LM})$ is the position of a given LM. Fitting parameters $ A $, $ \sigma $, and $ C $ have to be determined for each LM. A 15$\times$15 px$^2$ region was chosen to fit each peak centered at the position $(i_{LM},j_{LM})$, so $ i_{LM}-7 \leq i \leq i_{LM}+7 $ and $ j_{LM}-7 \leq j \leq j_{LM}+7 $. Such fitting region is optimal for our images  because it ensures that only one emission spot is included. The fitting parameters for each LM can be calculated via minimizing the following sum by least squares regression method
\begin{equation}\label{E:leastsq}
\sum_{i}\sum_{j} (p_{ij}-p_{ij}^e)^2,
\end{equation}
where $ p_{ij} $ is the original pixel value, $ i $ and $ j $ are the indices running over the fit region of each LM.

Obviously, most of the bright pixels will fall within one standard deviation of the mean. To classify a pixel $a$, located at $(i_a,j_a)$ within a fit region centered at a LM at $(i_{LM},j_{LM})$, as the $emission$ $pixel$ which contributes toward the emission area calculation, the following condition must be satisfied

\begin{equation}\label{E:emssionpixel}
(i_a-i_{LM})^2+(j_a-j_{LM})^2<\sigma^2.
\end{equation}
Alone, this condition is not enough to classify a pixel as the emission pixel. For example, consider the case when a background pixel is incorrectly detected and listed as a LM (we call it a false LM in Sec.~\ref{S:DB}). Gaussian fit of the region around such a false LM will yield a large standard deviation $\sigma$, but small amplitude $A$. This allows for filtering out fake LMs by applying threshold values for $A$ and $\sigma$. Put mathematically, LMs which satisfy the condition
\begin{equation}\label{E:fitcondition}
(\sigma>\sigma_{th})\lor(A<A_{th})
\end{equation}
are not real LMs and are discarded from the master LM list. Thus, they do not contribute to the local maxima count and the emission area calculation.

For the micrographs presented in this work, $\sigma_{th}=7$ and $A_{th}=3$ are chosen. For cases when a camera noise is high or emission spots are large, $A_{th}$ and $\sigma_{th}$ should be increased accordingly to produce reliable results. In Figs.~\ref{F:uncdimg}C and \ref{F:cntimg}C, the blue pixels identified as the emission pixels are overlaid with the emission spots shown in Figs.~\ref{F:uncdimg}A and \ref{F:cntimg}A, respectively. If the physical area of the image is known (we typically work with square images with a side length equal to the diameter of the cathode), then the apparent emission area is given by

\begin{equation}\label{E:earea}
\begin{split}
\text{Emission area (mm$^2$)} &= \text{Image area (mm$^2$)}\\
&\quad\times\frac{\text{Number of emission pixels}}{\text{Total number of pixels}}.
\end{split}
\end{equation}

Both Figs.~\ref{F:uncdimg}A and \ref{F:cntimg}A have 450$\times$450=202,500 pixels. There are 1,929 emission pixels in Fig.~\ref{F:uncdimg}C and 349 emission pixels in Fig.~\ref{F:cntimg}C. The displayed portion of a YAG:Ce screen in Figs.~\ref{F:uncdimg}A and \ref{F:cntimg}A have the area of 4.4$\times$4.4 mm$^2$ and 12.4$\times$12.4 mm$^2$, respectively. Then applying Eq.~\ref{E:earea}, the apparent emission area is 0.814 mm$^2$ for UNCD (Fig.~\ref{F:uncdimg}) and 0.264 mm$^2$ for CNTs (Fig.~\ref{F:cntimg}), while the total cathode surface area available for emission is 15.2 mm$^2$ for UNCD and 4.71 mm$^2$ for CNTs.

\section{Micrographs with Intense Background }\label{S:intense}

\begin{figure*}
	\centering\includegraphics[scale=0.23]{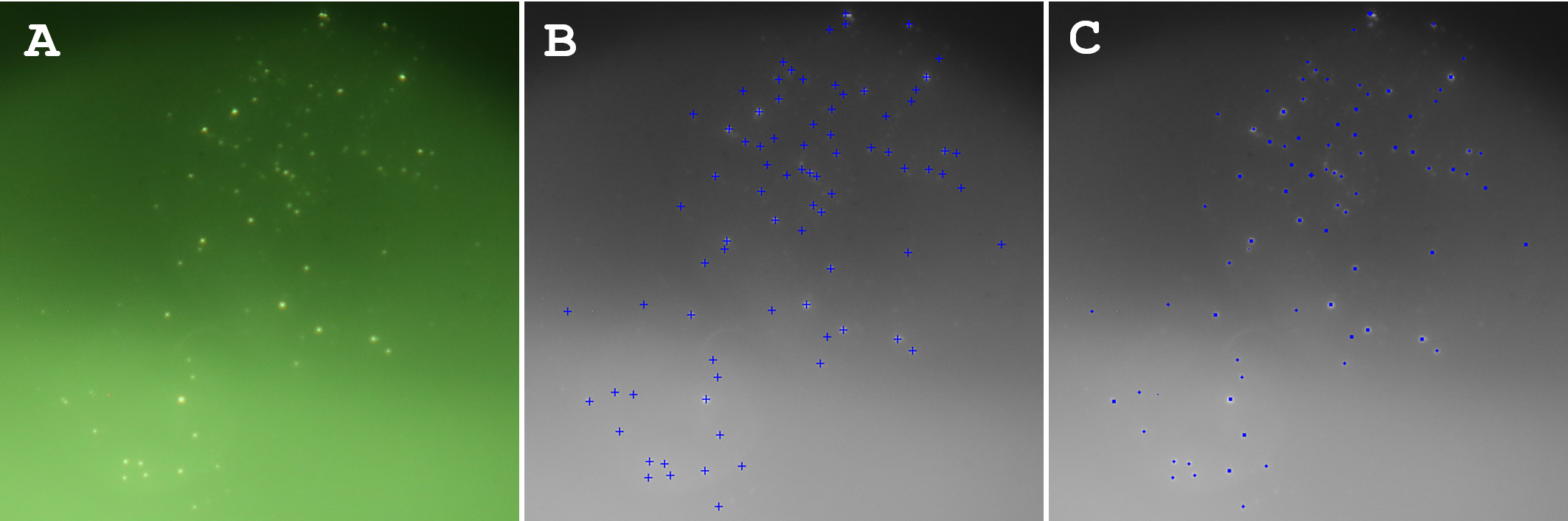}
		\caption{A. A typical micrograph obtained from a CNT fiber under dc field with glowing high-gradient background. The emission domains appear as bright peaks on the glow. Exact source of the glow is unknown. B. Detected LMs (shown with blue plus signs) overlaid with emission spots shown on the micrograph. C. Emission pixels (shown in blue), representing the projected emission area, overlaid with emission spots shown on the micrograph.}\label{F:cnt8}
\end{figure*}

\begin{figure*}
	\centering\includegraphics[scale=0.23]{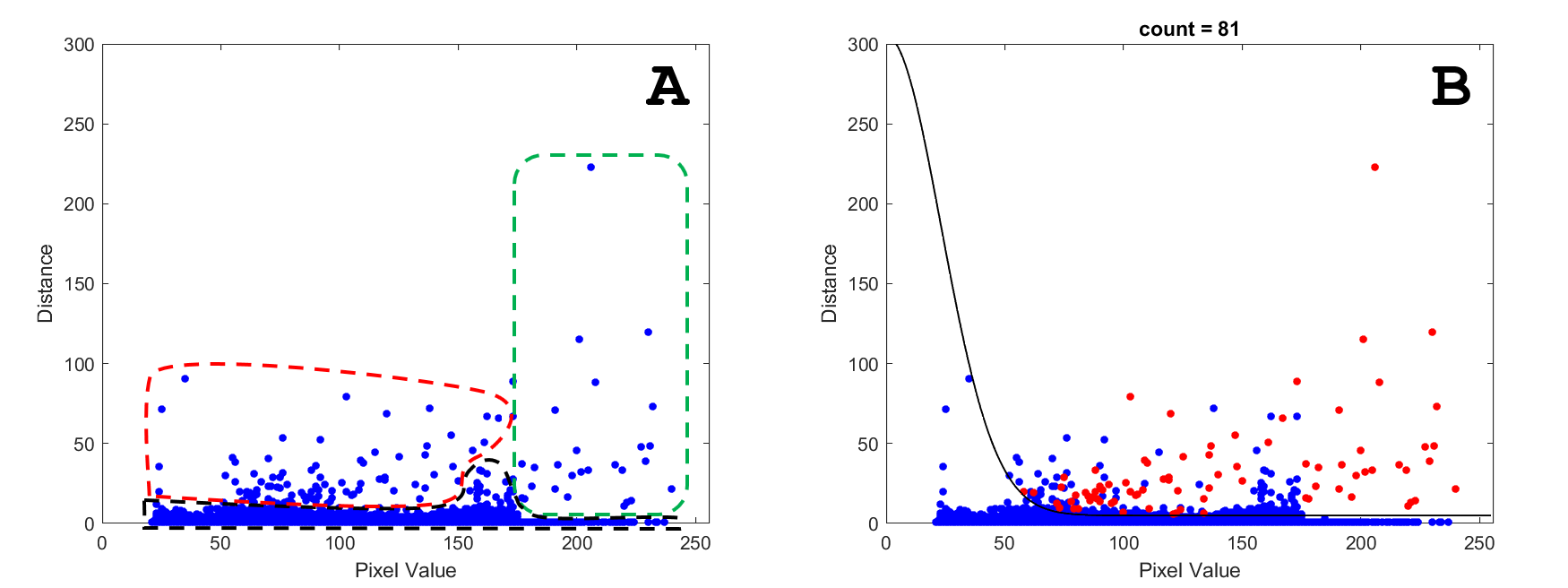}
\caption{A. Decision plot of extracted features for the micrograph obtained from a CNT fiber under the dc field with glowing high-gradient background. Unlike the previous cases, the background does not form a cluster. Instead, it is distributed over a wide pixel range shown in a black dashed region. Green dashed region consists of well separated LMs. Red region includes both background pixels and LMs, so the boundary should be drawn so that that region includes candidate LMs. Any false LMs are to be filtered out by Gaussian surface fitting. B. A black curve shows the applied decision boundary. All points above the curve are candidate LMs. Red points label finalized LM list after false LMs are filtered out by the surface fitting.}\label{F:cnt8db}
\end{figure*}

Micrographs with intense and glowing background are extremely challenging to process and analyze. One such example is shown in Fig.~\ref{F:cnt8}A. In this micrograph, the global background changes from dark at the top to very bright at the bottom, where the detection of emission spots becomes less effective. A corresponding decision plot is shown in Fig.~\ref{F:cnt8db}A. Unlike previous cases (Figs.~\ref{F:uncddb} and \ref{F:cntdb}), the background pixels do not cluster near the origin, but instead are distributed over a wide range of pixel values (shown by a black dashed region in Fig.~\ref{F:cnt8db}A). There are well separated bright pixels, corresponding to LMs, shown by a green dashed region. Because some of the glowing background pixels have values comparable or higher than those of some of the emission pixels, there is also a large region, shown by a red dashed line, where LMs and background pixels are intermixed.

It is important to note that any LM, excluded during the decision-making procedure, cannot be recovered during the surface-fitting procedure and calculating the emission area. However, any background pixel, identified and included as a LM, can be filtered out by the Gaussian surface fitting procedure. Therefore, the decision boundary should be $soft$, i.e. it should include all possible LM candidates (as opposed to a $rigid$ boundary that would exclude as many background pixels as possible). According to this approach, the decision boundary should be drawn on the decision plot (Fig.~\ref{F:cnt8db}) in such a way that the pixels from a red region are considered as LM candidates. Any false LM will be likely filtered out. One major drawback of applying the soft boundary is the increased computation time due to increased number of Gaussian fitting steps required for increased number of LMs. With the decision boundary applied in Fig.~\ref{F:cnt8db}B, all pixels above the black curve are LM candidates. Fig.~\ref{F:cnt8db}B highlights finalized array of LMs (after most or all false LMs were removed) in red. Same points are overlaid on the original image as blue crosses in Fig.~\ref{F:cnt8}B. Nearly perfect agreement can be seen upon visual inspection. Finally, Fig.~\ref{F:cnt8}C shows in blue the calculated apparent emission area. One can see how effectively the entire pattern recognition workflow performs even for images highly distorted by the background. The calculated number of emission pixels is 533 out of 202,500 pixels. The field of view in Fig.~\ref{F:cnt8}A is 11.2$\times$11.2 mm$^2$ yielding an emission area of 0.33 mm$^2$ while the total cathode area is 15.2 mm$^2$.

\section{Micrographs Under RF Field}\label{S:UNCD}

\begin{figure*}
	\centering\includegraphics[scale=0.23]{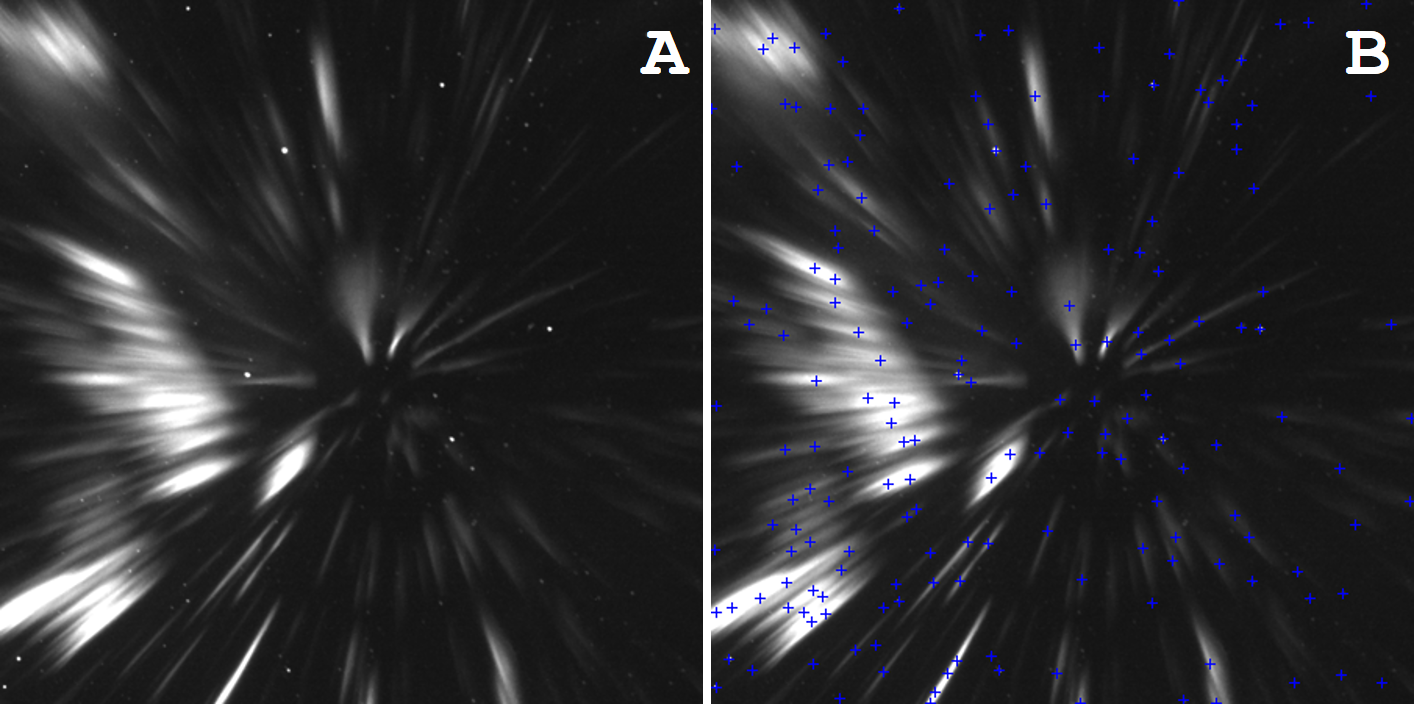}
	\caption{A) A typical micrograph obtained from an UNCD film under the rf field. Although physical emission spots are circular in shape on the cathode, they appear as streaks on the screen as a result of phase shift. B) Detected emission spots are shown with blue plus signs.}\label{F:uncdrfimg}
\end{figure*}

\begin{figure*}
	\centering\includegraphics[scale=0.23]{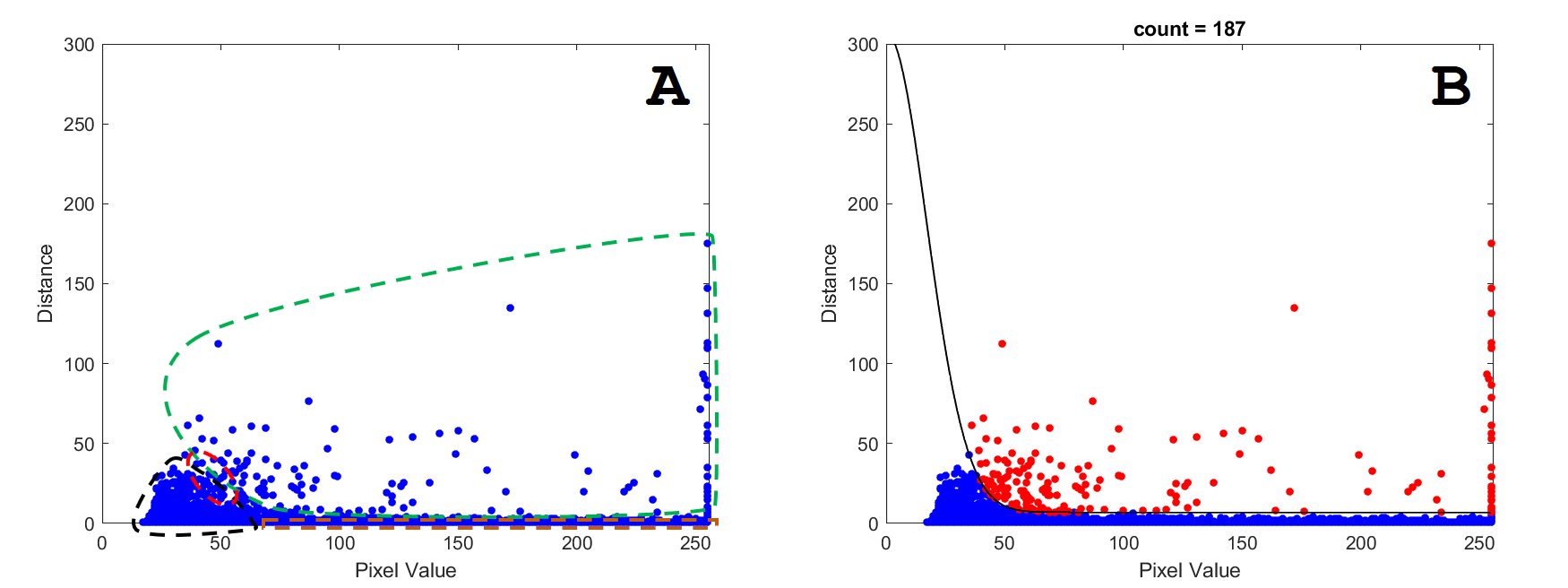}
	\caption{A) Decision plot of extracted features for the micrograph obtained from an UNCD under the rf field. Green, black, and brown dashed regions show where LMs, background pixels, and streak pixels are clustered. Dashed red region shows overlap, which causes some error. B) Black curve shows the applied Gaussian decision boundary. Red points indicate pixels classified as LMs.}\label{F:uncdrfdb}
\end{figure*}

As shown in Fig.~\ref{F:uncdrfimg}A, micro-emission zones, obtained from a field emitting planar UNCD operated under an rf field of a few 10's of MV/m, appear as bright stretched ellipses, but not circular spots. The emission spots are stretched along rays, which start at the center of a cathode and go in all directions. They vary in length and brightness and are non-uniformly distributed in polar coordinates. Unlike in dc, electrons are generated by and interact with the rf/microwave drive cycle in a wide phase window. The extended interaction phase window causes the ellipses (or streaks) to form (in contrast to circular spots typical for dc field). These streaks are essentially rotated projections of longitudinally stretched electron beamlets arriving from the cathode surface. From simulations\cite{Jiahang1,Jiahang2}, it is known that each streak corresponds to a physical emitter on the cathode surface. Therefore, counting streaks should be representative of the number of emitters and emission area dynamics against external power in the rf injector. At the same time, calculated apparent emission area will not be accurate due to rf phase effect distortion. Thus, the streaks can be counted through counting LMs.

The methodology given in Sec.~\ref{S:extraction} and \ref{S:DB} can also be used in this case. A circular kernel searches across the image for LMs. Diameter of the kernel has to be chosen such that it is smaller than the thickness of a streak. Kernel radius of 10 px was used for the 450$\times$450 px$^2$ image. A resulting decision plot is given in Fig.~\ref{F:uncdrfdb}A. Most of the LMs are bright and distant features (green dashed region), while background pixels are faint and are in close proximity (black dashed region). A very small portion of faint streaks atop a background of similar intensity is highlighted with red dashed loop. The Gaussian decision boundary, given in Eq.~\ref{E:gauss}, appears optimum for rf as well. One downside is that decision boundary needs to be rigid as the Gaussian fitting procedure (Sec.~\ref{S:area}) cannot be used. That means, once the decision boundary is applied, any false LM will not be filtered out. Therefore, the rf case is expected to be more prone to error. Nevertheless, the algorithm performed quite well. 

The distance between streaks is very diverse across the image. Some of streaks are too close to each other (may lead to undercounting). Some neighbor streaks are also overlapping. Their LMs will be bright and will have short distance features on a decision plot. The offset parameter $ k $ in Eq.~\ref{E:gauss} should be small so that all overlapping streaks are detected. On the other hand, these streaks are much longer and have almost uniform intensity over their length. Thus more than one LM per streak can be identified in a bright and short distance region of the decision plot (may lead to overcounting). The $ k $ should be large to count these streaks only once. Therefore, it should be chosen carefully to optimize this trade off. We found $ k=7 $ was an optimal choice for our rf images. The decision boundary shown with a black curve in Fig.~\ref{F:cnt8db}B was applied. Points classified as LMs (which are all points above the boundary because there is no the error correction procedure by curve fitting for rf images) are shown in red in Fig.~\ref{F:cnt8db}B. The finalized map of LMs counted for the image in Fig.~\ref{F:uncdrfimg}A is given in Fig.~\ref{F:uncdrfimg}B as an overlaid image. A fairly high accuracy was achieved.

\section{Conclusion}

In conclusion, a general pattern recognition algorithm for the analysis of electron emission micrographs was developed. Application of this algorithm was demonstrated and emphasized for large-area field emission cathodes by counting the number of field emission sites and calculating the apparent field emission area. It was demonstrated that the algorithm is applicable for analysis of micrographs obtained from field emission cathodes operated in both dc and rf environments where emission sites appear differently, either as circular- or streak-like features, respectively.

The algorithm is fast. Run on a personal laptop with Intel i5 2.5 GHz core, the LM counting and emission area calculation took 23.8 seconds for Fig.~\ref{F:uncdimg}, 18.9 seconds for Fig.~\ref{F:cntimg}, and 26.3 seconds for Fig.~\ref{F:cnt8}. The LM counting in Fig.~\ref{F:uncdrfimg} took 9.5 seconds (about half time as no emission area was calculated). It took the longest processing time for Fig.~\ref{F:cnt8} because a soft decision boundary was used and false LMs had to be sorted out.

Future work is being focused on unsupervised machine learning, especially one-class support vector machines to replace the Gaussian decision boundary and thus completely automate pattern recognition.

\section*{Acknowledgments}
This material is based upon work supported by the U.S. Department of Energy, Office of Science, Office of High Energy Physics under Award No. DE-SC0020429.

\bibliography{references}

\end{document}